\documentclass[a4paper,fleqn,10pt]{scrartcl}
\pdfoutput=1
\usepackage{authblk}
\usepackage{booktabs}
\usepackage{todonotes}
\presetkeys{todonotes}{inline}{}
\usepackage{amsmath}
\usepackage{amssymb}
\usepackage{commath}
\usepackage{array}
\usepackage{calc}
\usepackage{longtable}
\usepackage{multirow}
\usepackage{physics}
\usepackage{tensor}
\usepackage{upgreek}
\usepackage{graphicx}
\graphicspath{{figures/}}
\usepackage{xspace}
\usepackage{listings}
\usepackage[section]{placeins}
\usepackage{
  pgf,
  tikz}
\usetikzlibrary{
  patterns,
  shapes.multipart,
  arrows,
  trees,
  scopes,
  decorations.pathreplacing,
  decorations.pathmorphing,
  decorations.markings,
  decorations.text,
  positioning,
  calc
}

\usepackage[utf8]{inputenc}
\usepackage{mciteplus}
\usepackage[a4paper,pdfborder={0 0 0}]{hyperref}
\usepackage[format=hang,labelfont=bf,hypcap=true]{caption}
\usepackage{subfig}
\usepackage{sectsty}
\usepackage{relsize}

\allsectionsfont{\sffamily}
\subsubsectionfont{\mdseries\itshape\large}

\let\spreprint\empty
\newcommand{\preprint}[1]{\def\spreprint{\protect#1}}
\let\sinstitute\empty

\makeatletter
\renewcommand{\maketitle}{\begingroup
  \null\thispagestyle{empty}%
    \ifx\spreprint\empty
      \vskip 5ex
    \else
      \flushright\large\spreprint\vskip 2ex
    \fi
    \vskip 5ex
    \flushleft
      {\sffamily\bfseries\huge\@title}\vskip 2ex
      \@author\vskip 2ex
      \ifx\sinstitute\empty
      \else
        {\small\sinstitute}
      \fi
    \vskip 5ex
  \endgroup
}
\makeatother
\renewenvironment{abstract}{\begin{center}
  {\large\sffamily\bfseries Abstract: }
  \begin{minipage}[t]{0.75\textwidth}
}{\end{minipage}\end{center}\vskip 10ex}

\newcommand{\Sherpa}{S\protect\scalebox{0.8}{HERPA}\xspace}
\newcommand{\eerad}{E\protect\scalebox{0.8}{ERAD}3\xspace}

\newcommand{\Caesar}{C\protect\scalebox{0.8}{AESAR}\xspace}
\newcommand{\Ares}{A\protect\scalebox{0.8}{RES}\xspace}

\newcommand{\zcut}{\ensuremath{z_{\text{cut}}}}
\newcommand{\alphaS}{\alpha_\text{s}\xspace}
\newcommand{\muR}{\mu_\text{R}\xspace}
\newcommand{\LO}{\text{LO}\xspace}
\newcommand{\NLO}{\text{NLO}\xspace}    
\newcommand{\NLL}{\text{NLL}\xspace}
\newcommand{\NLLp}{\ensuremath{\text{NLL}^\prime}\xspace}
\newcommand{\NLOpNLL}{\ensuremath{\NLO+\NLL}\xspace}
\newcommand{\LOpNLL}{\ensuremath{\LO+\NLL}\xspace}    
\newcommand{\NLOpNLLp}{\ensuremath{\NLOpNLL^\prime}\xspace}
\newcommand{\LOpNLLp}{\ensuremath{\LOpNLL^\prime}\xspace}

\hypersetup{
  pdfauthor={Aude Gehrmann-De Ridder, Christian T Preuss, Daniel Reichelt,
             Steffen Schumann},
  pdftitle={Resummed predictions for three-jet event shapes in hadronic Higgs decays}
}

\preprint{MCNET-24-03\\ ZU-TH 16/24 \\IPPP/24/09}

\author[1,2]{Aude Gehrmann-De Ridder}
\author[3]{Christian T Preuss}
\author[4]{Daniel Reichelt}
\author[5]{Steffen Schumann}

\affil[1]{Institute for Theoretical Physics, ETH, CH-8093 Z{\"u}rich, Switzerland}
\affil[2]{Department of Physics, University of Z{\"u}rich, CH-8057 Z{\"u}rich, Switzerland}
\affil[3]{Department of Physics, University of Wuppertal, Gaussstr. 20, D-42119 Wuppertal, Germany}
\affil[4]{Institute for Particle Physics Phenomenology, Durham University, South Road, Durham DH1 3LE, U.K.}
\affil[5]{Institut für Theoretische Physik, Georg-August-Universität Göttingen, Friedrich-Hund-Platz 1, D-37077 Göttingen, Germany}

\title{NLO+NLL' accurate predictions for three-jet event shapes in hadronic Higgs decays}

\newcommand{\Nc}{\ensuremath{N_{\text{C}}}}

\newcommand{\D}{\ensuremath{\mathrm{d}}}

\newcommand{\mods}[1]{\ensuremath{\left\vert#1 \right\vert}}
\newcommand{\FFun}{\ensuremath{\mathcal{F}}}
\newcommand{\gammaE}{\ensuremath{\gamma_\mathrm{E}}}
\DeclareUnicodeCharacter{2212}{-}

\begin{document}
\maketitle

\begin{abstract}
  We present resummed predictions at next-to-leading logarithmic accuracy
  matched to the exact next-to-leading order results for a set of classical
  event-shape observables in hadronic Higgs decays, \emph{i.e.}, for the
  channels $H\to gg$ and $H\to b\bar{b}$. We furthermore consider soft-drop
  grooming of the hadronic final states and derive corresponding \NLO+\NLLp
  predictions for the groomed thrust observable. Differences in the QCD radiation
  pattern of gluon- and quark-initiated final states are imprinted in the
  event-shape distributions, offering separation power for the two decay channels.
  In particular, we show that ungroomed event shapes in $H\to gg$ decays develop
  a considerably harder spectrum than in $H\to b\bar b$ decays.
  We highlight that soft-drop grooming can substantially alter this behaviour,
  unless rather inclusive grooming parameters are chosen.
\end{abstract}

\tableofcontents
\clearpage

\section{Introduction}

A new high-energy lepton collider, such as the FCC-ee~\cite{FCC:2018byv,FCC:2018evy}, the
CEPC~\cite{CEPCStudyGroup:2018ghi}, or the ILC~\cite{ILC:2013jhg}, operated in the Higgs-factory
mode presents a key scenario for future particle-physics collider experiments. The primary goal
of these facilities lies in the precise determination of the properties of the Higgs-boson particle,
recently discovered by the ATLAS and CMS experiments at the LHC~\cite{ATLAS:2012yve,CMS:2012qbp}.
This includes detailed measurements of its decay width and branching fractions, and hence its
couplings, as well as differential distributions of its decay final states at unprecedented
precision.

The dominant production mode of Higgs bosons at leptonic Higgs factories proceeds via the
Higgs-strahlungs process, \emph{i.e.}, in association with a $Z$-boson, through $e^+e^-\to ZH$.
The absence of hadronic initial-state radiation in lepton--lepton collisions facilitates the
investigation of hadronic Higgs-boson decays that are inaccessible at hadron colliders such
as the LHC. This is in particular the case for the Higgs decaying into gluons. 
To this end, leptonic decays of the gauge boson are considered, leading to the
experimental signature $e^+e^-\to l^+l^-+X_{\text{QCD}}$, with $X_{\text{QCD}}$ the hadronic
final state originating from the Higgs-boson decay, \emph{e.g.}, through $H\to q\bar{q}$
and $H\to gg$. For the cases $H\to b\bar{b}$ and $H\to c\bar{c}$, displaced
vertices emerging from weak decays of bottom- and charm-flavoured hadrons can be instrumented
~\cite{CMS:2018nsn,ATLAS:2020jwz,CMS:2019hve,ATLAS:2022ers}, providing means to disentangle
these from the gluon decay mode. However, also differences in the QCD radiation pattern for
quark- and gluon-initiated final states offer discriminatory power \cite{Gao:2016jcm,Gao:2019mlt,Luo:2019nig,Gao:2020vyx,Knobbe:2023njd,Wang:2023azz}.
A particular class of observables sensitive to
these radiative corrections are event-shape variables, that probe the geometric properties
of the hadronic final state. Event-shape observables played a prominent role in the analysis
of QCD final states produced in electron--positron annihilation at
LEP~\cite{ALEPH:1990iba,OPAL:1990xiz,L3:1992btq,DELPHI:1999vbd,DELPHI:2003yqh}. In particular, they
have been, and still are, instrumental for precision extractions of the strong coupling
$\alphaS$~\cite{dEnterria:2022hzv}.
These analyses pertaining to three-jet-like event shapes in hadronic $Z$-boson and off-shell photon decays
are facilitated by precision calculations, including the evaluation of next-to-next-to-leading order
(NNLO) corrections in the strong coupling
\cite{Gehrmann-DeRidder:2007vsv,Gehrmann-DeRidder:2009fgd,Gehrmann-DeRidder:2014hxk,Weinzierl:2009nz,Weinzierl:2009ms,Weinzierl:2009yz,DelDuca:2016ily,Kardos:2018kth}
and the resummation of logarithmically enhanced terms \cite{Becher:2011pf,Becher:2012qc,Balsiger:2019tne,Hoang:2014wka,Banfi:2014sua,Banfi:2018mcq,Bhattacharya:2022dtm,Bhattacharya:2023qet}.
Recently, also power corrections related to hadronisation effects have been taken into account \cite{Gehrmann:2010uax,Luisoni:2020efy,Caola:2021kzt,Caola:2022vea,Agarwal:2023fdk}.
Furthermore, measurements of event shapes provide a crucial test-bed for Monte Carlo event
generators, based on parton-shower simulations and phenomenological models for
hadronisation~\cite{Buckley:2011ms,Campbell:2022qmc}. It is worth noting, that QCD final states
produced at LEP are dominated by the hadronic decays $\gamma^*/Z\to q\bar{q}$, such that gluon-initiated
jets are much less constrained. This motivates dedicated studies of
event shapes in hadronic Higgs-boson decays, both at fixed order in perturbation theory and
for resummation calculations.

Beyond the direct comparison to existing or future experimental data,
theoretical predictions for Higgs-boson decays to quarks and in particular gluons
are used as proxies in jet-flavour tagging studies, in particular to investigate
related theoretical and modelling uncertainties. Given that a theoretically clean
definition of a quark/gluon jet is in itself a complicated research
topic~\cite{Banfi:2006hf,Komiske:2018vkc,Caletti:2021oor,Caletti:2022glq,
  Caletti:2022hnc,Czakon:2022wam,Gauld:2022lem,Caola:2023wpj,Andersen:2024czj},
it is often unclear where to start if one seeks for a detailed understanding
of the systematics involved. A common solution to study light-quark and
gluon tagging, for example in the studies performed in~\cite{Andersen:2016qtm, Gras:2017jty,
  Mo:2017gzp, Reichelt:2017hts}, is to take Higgs-boson decays to gluons as a clean
probe of gluon-induced jets, and the decay of a singlet to quarks as the corresponding
counterpart. A detailed understanding of the involved perturbative inputs, provided for
a large variety of relevant observables, is a crucial element for future studies in this area.

First resummed predictions for thrust in $H\to gg$ decays at $\text{NNLL}^\prime$ accuracy have been presented in~\cite{Mo:2017gzp}.
$\text{NNLL}^\prime$ results for the class of angularity variables have recently been derived in ~\cite{Zhu:2023oka,Yan:2023xsd}.
A few approaches that combine a fixed-order calculation matched to NNLL resummation with full event-generation frameworks
have been derived for hadronic Higgs decays. The \textsc{Geneva} collaboration compiled results based on the NNLO
fixed-order calculation for two-body decays with NNLL 2-jettiness resummation for both Higgs decay channels~\cite{Alioli:2020fzf}.
Similarly, the approach presented in \cite{Hu:2021rkt} utilises NNLL resummation of the thrust observable to combine two-body
decays at NNLO with parton showers. Within the \textsc{MiNLO} scheme, a NNLO calculation of the $H\to b\bar b$ decay is embedded
into a full event-simulation framework via NNLL resummation of the three-jet resolution variable in the Cambridge--Aachen
algorithm~\cite{Bizon:2019tfo}.

We here consider the matching of exact next-to-leading order QCD predictions for six classic
three-jet event shapes recently presented in Ref.~\cite{Coloretti:2022jcl} with resummed predictions
at next-to-leading logarithmic accuracy, evaluated using the implementation of the \Caesar
resummation formalism~\cite{Banfi:2004yd} in the \Sherpa event generator
framework~\cite{Gerwick:2014gya}. We thereby consider in particular thrust, $C$-parameter,
heavy-hemisphere mass, total and wide jet broadening, and the Durham three-jet resolution
variable $y^\mathrm{D}_{23}$. Furthermore, we here derive and present predictions for soft-drop
thrust~\cite{Marzani:2019evv} for three values of the soft-drop angular parameter $\beta$,
namely $\beta=0,1,2$.

Our paper is organised as follows: In Sec.~\ref{sec:setup} we describe our
theoretical framework, providing details about the fixed-order and resummed
computations. Furthermore, we introduce the considered event-shape observables.
In Sec.~\ref{sec:validation} we cross-validate our calculations by considering the
soft limit for the fixed-order predictions and the expansions of the
resummation to ${\cal{O}}(\alphaS)$ and ${\cal{O}}(\alphaS^2)$. Our final
matched predictions with \NLOpNLLp accuracy are presented in
Sec.~\ref{sec:resummed_predictions}. Finally, we give our conclusions and
a brief outlook on future research avenues in Sec.~\ref{sec:conclusions}.

\section{Theoretical framework}\label{sec:setup}

\begin{figure}[!t]
  \includegraphics[width=0.95\textwidth]{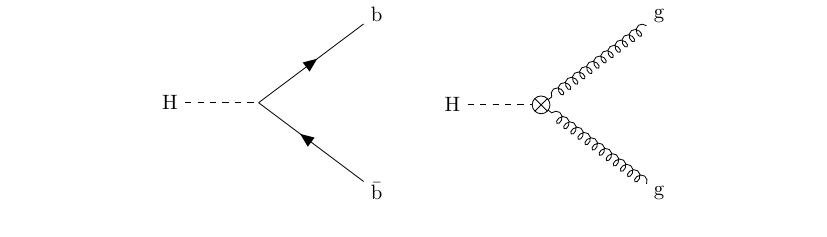}
  \caption{Higgs to two-parton contributions to the hadronic Higgs-boson decays:
    $H\to b\bar{b}$ proportional to the Yukawa coupling $y_b$, and
    $H\to gg$ mediated through an effective coupling.}\label{fig:Higgs_HadDecays}
\end{figure}

The computation of three-jet like event-shape observables related to hadronic Higgs decays is performed in an effective field theory 
framework, where two distinct classes of processes arising through the two-parton decay modes $H \to gg$ and $H \to b \bar{b}$ are considered.
In the former case, the Higgs boson couples to gluons via a heavy-quark loop which decouples in the limit of an infinitely-large top-quark mass.
More precisely, we compute differential decay rates in the limit of vanishing light-quark 
masses, while keeping a non-vanishing Yukawa coupling $y_b$ only for the bottom quark and an infinitely large top-quark mass.  As a consequence,
the hadronic decay modes of the Higgs boson can be divided into two categories at parton-level:
In the first category, the Higgs decays into a bottom--anti-bottom pair and is related to the non-vanishing Yukawa coupling $y_b$.
In the second category, related to the decay of the Higgs boson to gluons, observables are computed in an effective theory approach in which
the Higgs boson couples directly to gluons through an effective Higgs-gluon-gluon vertex.  
The latter vertex originates from a closed top-quark loop, considered in
the infinite top-mass limit~\cite{Ellis:1975ap,Shifman:1979eb,Kniehl:1995tn}.
Both Higgs decay modes to two partons are depicted in Fig.~\ref{fig:Higgs_HadDecays}, where the effective $Hgg$ vertex is represented as a crossed dot.

The framework used here can be formulated in terms of an effective Lagrangian as presented in \cite{Coloretti:2022jcl}:
\begin{equation}
    \mathcal{L}_\mathrm{Higgs} = -\frac{\lambda(m_t,\muR)}{4}HG_{\mu\nu}^aG^{a,\mu\nu} + \frac{y_b(\muR)}{\sqrt{2}}H\bar{\psi}_b\psi_b \, .
\label{eq:lagrangian}
\end{equation}
In this context, the effective $Hgg$ coupling is given in terms of the Wilson coefficient $C(m_t,\muR)$ by
\begin{equation}
    \lambda(m_t,\muR) = -\frac{\alphaS(\muR)C(m_t,\muR)}{3\uppi}\frac{4\uppi\alpha}{2m_\mathrm{W}\sin\theta_\mathrm{W}} \, ,
\end{equation}
and the $Hb\bar{b}$ Yukawa coupling reads
\begin{equation}
    y_b(\muR) = \overline{m}_b(\muR)\frac{4\uppi\alpha}{\sqrt{2}m_\mathrm{W}\sin\theta_\mathrm{W}} \, .
\end{equation}
Here, $\alpha$ denotes the electromagnetic coupling and $\theta_\mathrm{W}$ is the Weinberg angle.
Both $\lambda$ and $y_b$ are subject to renormalisation, which we perform at scale $\muR$ in the $\overline{\text{MS}}$ scheme using $N_\mathrm{F} = 5$. The top-quark Wilson coefficient is evaluated at first order in $\alphaS$ using the results of \cite{Inami:1982xt,Djouadi:1991tk,Chetyrkin:1997iv,Chetyrkin:1997un,Chetyrkin:2005ia,Schroder:2005hy,Baikov:2016tgj}.

It is important to stress that the terms in Eq.~\eqref{eq:lagrangian} do not interfere under the assumption of kinematically massless quarks. In particular, they do not mix under renormalisation \cite{Gao:2019mlt}. This allows us to define two separate Higgs-decay categories, as mentioned above, and to compute higher-order corrections independently for each. 
All partonic contributions with three hard final-state partons at Born level that are needed for the fixed-order parton-level calculations used in this paper have been presented up to $\mathcal{O}(\alphaS^2)$ in \cite{Coloretti:2022jcl}.

In what follows, we will describe the calculational tools and setups used to derive
our fixed-order and resummed predictions for a set of widely considered event-shape
observables obtained in both Higgs-decay categories. 
For the case of the thrust variable~\cite{Brandt:1964sa,Farhi:1977sg}
we furthermore consider its soft-drop groomed variant~\cite{Marzani:2019evv},
where the hadronic final state prior to the observable evaluation is subjected
to soft-drop grooming~\cite{Larkoski:2014wba}. 

\subsection*{NLO calculation}

For an infrared and collinear safe event-shape observable $V$, a function of the hadronic final state momenta, that yields
the observable value $v$, the differential hadronic-decay width can be written up to NLO in the strong coupling $\alphaS$ as
\begin{equation}\label{eq:fo_xsec}
  \frac{1}{\Gamma_{2j}}\frac{\D \Gamma}{\D v} = \frac{\alphaS(\muR)}{2\uppi}\frac{\D A}{\D v} + \left(\frac{\alphaS(\muR)}{2\uppi}\right)^2\frac{\D B}{\D v} \,, 
\end{equation}
with the leading-order (LO) coefficient $A$ and the next-to-leading order (NLO) coefficient $B$. In this context, $\Gamma_{2j}$ defines the inclusive partial decay width in the respective decay channel,
\begin{equation}
  \Gamma_{b\bar b}(\muR) = \frac{y_b^2(\muR)m_H\Nc}{8\uppi} \, , \quad \Gamma_{gg}(\muR) = \frac{\alphaS^2(\muR)G_\mathrm{F} m_H^3}{36\uppi^3\sqrt{2}} \, ,
\end{equation}
where $\Nc = 3$ is the number of colours, $G_\mathrm{F}=4\uppi\alpha/(4\sqrt{2}m_\mathrm{W}^2\sin^2\theta_\mathrm{W})$ denotes the Fermi constant and $m_H$ stands for the Higgs-boson mass.

We calculate the perturbative coefficients $A$ and $B$ using the \eerad parton-level event generation framework. Originally developed for NNLO calculations of
event shapes \cite{Gehrmann-DeRidder:2007vsv,Gehrmann-DeRidder:2009fgd} and jet distributions \cite{Gehrmann-DeRidder:2008qsl} for $\gamma^*/Z$-decays related to $e^+e^-\to 3j$, \eerad has
recently been extended to include hadronic Higgs decays to three \cite{Coloretti:2022jcl,Aveleira:2024dcx} and four jets \cite{Gehrmann-DeRidder:2023uld} in the $H\to b\bar{b}$
and $H\to gg$ channels at NLO level.
As alluded to above, the computation of three jet-like Higgs event-shapes involving the hadronic Higgs decays 
to three hard partons at the Born-level are performed assuming an effective Lagrangian obtained under the assumptions of an infinitely heavy top-quark, {\em i.e.}, $m_t \to \infty$,
and massless light quarks, $m_q \equiv 0$, where only the $b$-quark has a non-vanishing Yukawa coupling $y_b > 0$.

Infrared singularities are regulated using the antenna-subtraction scheme \cite{Campbell:1998nn,Gehrmann-DeRidder:2005btv,Currie:2013vh} to construct real and
virtual subtraction terms. All tree-level and virtual matrix elements are implemented in fully analytic form, assuring stable numerical predictions even for
infrared kinematics. All event-shape distributions are calculated above a cut of $v_\mathrm{cut}$, so that $v > v_\mathrm{cut}$.
For all observables except the three-jet resolution scale, a cut of $v_\mathrm{cut} = 10^{-5}$ is chosen; for the three-jet resolution variable, the cut is lowered to $v_\mathrm{cut} = 10^{-10}$.
For the renormalisation scale entering the calculation we use as central value $\muR=m_H$. We employ one- and two-loop running for the strong coupling
$\alphaS$ for the LO and NLO calculations, respectively. As input at scale $m_Z$ we use $\alphaS(m_Z)=0.118$.

\subsection*{NLL resummation}

We derive all-orders results for the event-shape observables at next-to-leading logarithmic (NLL)
accuracy using the implementation of the \Caesar formalism~\cite{Banfi:2004yd} in the \Sherpa event-generator
framework~\cite{Gleisberg:2008ta,Bothmann:2019yzt}, first presented in~\cite{Gerwick:2014gya}. Within
the \Caesar approach one considers infrared- and collinear-safe (IRC-safe) observables $V$ that scale for the
emission of a soft-gluon of relative transverse momentum $k_t^{(l)}$, rapidity $\eta^{(l)}$,
and azimuthal angle $\phi^{(l)}$ with respect to leg $l$ of the Born-level process as
\begin{equation}\label{eq:CAESAR_param}
  V(k)=\left(\frac{k_{t}^{\left(l\right)}}{\mu_Q}\right)^{a}e^{-b_{l}\eta^{\left(l\right)}}d_{l}\left(\mu_Q\right)g_{l}\left(\phi^{(l)}\right)\,.
\end{equation}

For the global event-shape observables considered here and a specific partonic channel $\delta$, the
NLL accurate all-order cumulative cross section for observable values up to $v$, with $L=-\ln(v)$, can be written as 
\begin{equation}\label{eq:CAESAR_master}
  \Gamma_\text{NLL}(v)
   = \int \mathop{\D\mathcal{B}}
    \frac{\mathop{\D\Gamma_{2j}}}{\mathop{\D\mathcal{B}}}
    \exp\left[-\sum_{l\in\mathcal{B}}R_l(L)\right]
    \mathcal{F}(L)\,.
\end{equation}
Here, $\mathop{\D\Gamma_{2j}}/\mathop{\D\mathcal{B}}$ denotes the Born-level cross section;
$R_l$ the collinear radiators; and $\mathcal{F}$ is the multiple-emission function.

In Ref.~\cite{Baron:2020xoi} the \Caesar formalism, and in particular the radiator functions $R_l$,
and the corresponding implementation in \Sherpa were extended to include the phase-space constraints
given by soft-drop grooming with general parameters \zcut\ and $\beta$~\cite{Larkoski:2014wba}. The framework
has recently been used to obtain resummed predictions for soft-drop thrust~\cite{Marzani:2019evv} and
multijet resolution scales~\cite{Baberuxki:2019ifp} in electron--positron annihilation, as well as \NLOpNLLp
predictions for soft-drop groomed hadronic event shapes~\cite{Baron:2020xoi}, jet angularities in dijet and
$Z$+jet production at the LHC~\cite{Caletti:2021oor, Caletti:2021ysv,
  Reichelt:2021svh} as well as RHIC~\cite{Chien:2024uax} and, lately, for plain
and groomed 1-jettiness in deep inelastic scattering~\cite{Knobbe:2023ehi, H1:2024pvu, H1:2024aze}.

\subsection*{$\text{NLO}+\text{NLL}^\prime$ matching}\label{sec:matching}

We here consider combining the resummation to NLL with exact NLO results for
three-jet event shapes using a multiplicative matching scheme, thereby aiming for $\NLOpNLLp$ accuracy.
Starting from Eq.~\eqref{eq:fo_xsec}, we can introduce the decay width for events with observable values
up to $v$,
\begin{equation}
  \Gamma(v) \equiv \int_0^v \D v \frac{\D\Gamma_{2j}}{\D v}  = \Gamma_{2j}
  \left(1+\frac{\alphaS(\muR)}{2\uppi} A(v) + \left(\frac{\alphaS(\muR)}{2\uppi}\right)^2 B(v)\right)\,,
\end{equation}
where $A(v)$ is normalised so that the full integral over $v$, \emph{i.e.}, $A(v_\text{max})$,
reproduces the NLO correction to $\Gamma_{2j}$.
Note that this does not change the functional form of Eq.~\eqref{eq:fo_xsec}.

The logarithmic structure of $\Gamma(v)$ at order $\alphaS^n$
contains up to $2n$ powers of the logarithm $L = -\ln(v)$. Hence, we separate
each of $A(v)$, $B(v)$ into terms $A_i$, $B_j$ containing $i\leq2$ and $j\leq4$
powers of $L$, and a finite remainder that vanishes in the $v\to0$
limit. Explicitly, we write
\begin{align}
  A(v) &= L^2 A_2 + L A_1 + A_0 + A_\text{finite}(v) \label{eq:A_logs}\,, \\
  B(v) &= L^4 B_4 + L^3 B_3 + L^2 (B_2 + A_0 A_2) + L B_1 + B_0 + B_\text{finite}(v)\,.\label{eq:B_logs}
\end{align}

We have chosen to make it explicit in our notation that the coefficient of $L^2$
in $B(v)$ receives contributions that can be constructed from the ones present
in $A(v)$ as well as generic NLO corrections $B_2$. In principle, the same is true for
$B_1$, $B_0$, and $B_\text{finite}$, which we suppress since these coefficients are
not completely restored at NLL$^\prime$ accuracy anyway. Comparing to other
conventions used in the literature, for example \cite{Catani:1992ua}, we can
identify the constants and finite remainders with
\begin{align}
  A_0=\mathcal{C}_1\,,\;\;& A_\text{finite}(v)=\mathcal{D}_1(v)\,,\\
  B_0=\mathcal{C}_2\,,\;\;& B_\text{finite}(v)=\mathcal{D}_2(v)\,.
\end{align}

For observables satisfying the conditions for automated NLL resummation outlined in
\cite{Banfi:2004yd}, in particular recursive IRC safety, the NLL resummed cross
section takes the well-known form
\begin{equation}
  \Gamma_\text{NLL}(v) = \Gamma_{2j}\exp(L g_1(\alphaS L) + g_2(\alphaS L))\,.
\end{equation}
We follow the notion of Ref.~\cite{Banfi:2004yd} and define the logarithmic accuracy by the
terms appearing in the exponent as opposed to its expansion, \emph{i.e.}, we refer
to the cross section containing $g_1$ as LL accurate and after including $g_2$ NLL
accurate, while including the next term $\alphaS g_3(\alphaS L)$ would
correspond to NNLL accuracy. We denote by \NLLp (and corresponding for higher
accuracy) expressions that have additionally the mixed contributions arising from the
$\mathcal{C}_1$ term included.
When expanding the resummed all-orders cross section to second order in $\alphaS$,
and sorting according to logarithmic enhancement in the same terms as in Eqs.~\eqref{eq:A_logs},
\eqref{eq:B_logs} we obtain

\begin{equation}\label{eq:GammaNLL_expansion}
  \Gamma_\text{NLL}(v) \sim \Gamma_{2j}\left(1+\frac{\alphaS(\muR)}{2\uppi} A_\text{NLL} +
  \left(\frac{\alphaS(\muR)}{2\uppi}\right)^2 B_\text{NLL} + \mathcal{O}(\alphaS^3)\right)\,,
\end{equation}
where
\begin{eqnarray}
  A_\text{NLL}(v) &=& L^2 A_2 + L A_1\,, \label{eq:ANLL_logs}\\
  B_\text{NLL}(v) &=& L^4 B_4 + L^3 B_3 + L^2 B_2\,.
\end{eqnarray}
We can then write the expression for a multiplicative matching up to order $\alphaS^2$ as
\begin{align}\label{eq:GammaMatch}
  \Gamma(v) = \Gamma_\text{NLL}(v)
  \Bigg[1 &+ \frac{\alphaS(\muR)}{2\uppi}\Bigg(A(v)-A_\text{NLL}(v)\Bigg) \\
  &+\left(\frac{\alphaS(\muR)}{2\uppi}\right)^2 \Bigg(B(v)-B_\text{NLL}(v)-A_\text{NLL}(v)\Big(A(v)-A_\text{NLL}(v)\Big)\Bigg)\Bigg]\,.\nonumber
\end{align}
As we analyse the finite remainder after subtracting the logarithmic terms
\begin{equation}
  A_\text{remain}(v) \equiv \frac{\alphaS}{2\uppi}A(v)-\frac{\alphaS}{2\uppi}A_\text{NLL}(v)
\end{equation}
we can read off from Eqs.~\eqref{eq:A_logs} and \eqref{eq:ANLL_logs}
\begin{equation}\label{eq:Aremain}
   A_\text{remain}(v) = \frac{\alphaS}{2\uppi}\left(A_0 + A_\text{finite}(v)\right) \to \frac{\alphaS}{2\uppi}A_0 = \frac{\alphaS}{2\uppi}\mathcal{C}_1
  \;\text{as}\; v\to0\,,
\end{equation}
such that the exact ${\cal{O}}(\alphaS)$ coefficient in the small-observable limit is recovered.
Naively, the NLO remainder defined as 
\begin{equation}\label{eq:Bremain}
  B_\text{remain}(v) \equiv \left(\frac{\alphaS}{2\uppi}\right)^2B(v)-
  \left(\frac{\alphaS}{2\uppi}\right)^2B_\text{NLL}(v) \sim
  \alphaS^2L^2 A_2 A_0 +
  \mathcal{O}\left(\alphaS^2 L,v\right)\,,
\end{equation}
still contains a $L^2$ coefficient, as well as NNLL-type contributions of order $\alphaS^2L$, and terms
suppressed by positive powers of $v$. We can reproduce this expression effectively by defining
\begin{equation}
 B^\prime_\text{NLL}(v) = B_\text{NLL}(v)+A_\text{NLL}(v)\left(\frac{2\uppi}{\alphaS}\right)A_\text{remain}(v) = L^4 B_4 + L^3 B_3 +
L^2 B_2 + L^2 A_2 A_0 + \mathcal{O}\left(L,v\right)\,,
\end{equation}
such that the remainder given by
\begin{equation}\label{eq:Bpremain}
  B^\prime_\text{remain} \equiv \left(\frac{\alphaS}{2\uppi}\right)^2 B(v)-
  \left(\frac{\alphaS}{2\uppi}\right)^2 B^\prime_\text{NLL}(v)
\end{equation}
has all leading and next-to-leading logarithms properly subtracted. Accordingly,
Eq.~\eqref{eq:GammaMatch} indeed achieves \NLOpNLLp accuracy for the observable
distribution.

\subsection*{Observable definitions}

As alluded to above, we here want to study a set of widely considered
event-shape observables that analyse the momentum distribution of the
hadronic final-state particles, offering potential to discriminate quark-
from gluon-initiated final states. In practice, we consider all decay
products of the Higgs boson in its rest frame, such that in what follows
we identify $s=m^2_H$. 

\paragraph{3-Jet Resolution}
The three-jet resolution variable in a given jet algorithm is calculated as the minimal distance $y_{23} := \min_{(i,j)}\{y_{ij}\}$ among all pairs $(i,j)$ in a three-jet configuration. It determines the value of the distance measure for the transition from a three-jet to a two-jet event.
Here, we consider the Durham jet algorithm with the particle-distance measure \cite{Catani:1991hj,Brown:1990nm,Brown:1991hx,Stirling:1991ds,Bethke:1991wk}
\begin{equation}
  y_{ij}^{\mathrm{D}} = \frac{2\min (E_i^2,E_j^2)(1-\cos\theta_{ij})}{s} \, .
\end{equation}
We employ the so-called ``E-scheme'', in which in each step of the algorithm the pair with smallest resolution $y_{ij}^{\mathrm{D}}$ is replaced by a pseudo-jet with four-momentum equal to the sum of the four-momenta of $i$ and $j$.

The \Caesar parameters of the Durham three-jet resolution are $a=2$ and $b=0$. We calculate the remainder function $\FFun$ appearing in Eq.~\eqref{eq:CAESAR_master} numerically using the algorithm of \cite{Baberuxki:2019ifp}.

\paragraph{$C$-Parameter}
The $C$-parameter,
\begin{equation}
  C = 3(\lambda_1\lambda_2 + \lambda_2\lambda_3 + \lambda_3\lambda_1) \, ,
\end{equation}
is defined in terms of the three eigenvalues $\lambda_{1,2,3}$ of the linearised momentum tensor \cite{Parisi:1978eg,Donoghue:1979vi}
\begin{equation}
  \Theta^{\alpha\beta} = \frac{1}{\sum\limits_j\mods{\vec{p}_j}}\sum\limits_i\frac{p_i^\alpha p_i^\beta}{\mods{\vec{p}_i}} \, ,\;\; \text{where } \alpha,\beta \in \{1,2,3\} \,,
\end{equation}
such that $0\leq \lambda_{1,2,3}\leq 1$, $\sum_i\lambda_i=1$ and consequently $0\leq C\leq 1$.

The \Caesar parametrisation for the $C$-parameter is $a=1$, $b=1$ and the remainder function appearing in Eq.~\eqref{eq:CAESAR_master} is the standard one for additive observables \cite{Catani:1992ua,Catani:1991kz,Banfi:2001bz}
\begin{equation}
  \FFun_\mathrm{add}(L) = \frac{\exp\left(-\gammaE R'(L)\right)}{\Gamma(1+R'(L))} \,,
  \label{eq:FFunctionAdditive}
\end{equation}
with $R'(L)=\partial R/\partial L=\partial_L R$ and $\gammaE$ the Euler--Mascheroni constant.

\paragraph{Thrust}
For multi-jet final states from colour singlet decays, thrust is defined as \cite{Brandt:1964sa,Farhi:1977sg}
\begin{equation}
  T = \max\limits_{\vec{n}}\left(\frac{\sum\limits_i \mods{\vec{p}_i\cdot\vec{n}}}{\sum\limits_i \mods{\vec{p}_i}} \right) \, ,
\end{equation}
where the sum of three-momenta is maximised over the direction of $\vec{n}$. The so-called thrust axis is defined by the unit vector $\vec{n}_T$ which maximises the expression on the right-hand side. For two-jet events, the thrust approaches unity, $T\to 1$, while for three-particle events it holds $T \geq \frac{2}{3}$. For practical purposes, we consider the related observable
\begin{equation}
  \tau \equiv 1-T = \min\limits_{\vec{n}}\left(1-\frac{\sum\limits_i \mods{\vec{p}_i\cdot\vec{n}}}{\sum\limits_i \mods{\vec{p}_i}} \right) \, ,
\end{equation}
so that $\tau > 0$ measures the departure from the two-particle limit.

The \Caesar parameters of thrust according to Eq.~\eqref{eq:CAESAR_param} read $a=1$ and $b=1$. As thrust
is an additive observable, its ${\cal{F}}$ function entering Eq.~\eqref{eq:CAESAR_master} is again given by Eq.~\eqref{eq:FFunctionAdditive}.

\paragraph{Heavy-jet mass}
We define the hemispheres $\mathcal{H}_\mathrm{L}$ and $\mathcal{H}_\mathrm{R}$ so that they are separated by a plane orthogonal to the thrust axis. The heavy-jet mass is then calculated as \cite{Clavelli:1981yh}
\begin{equation}
  \rho_\mathrm{H} \equiv \frac{M_\mathrm{H}^2}{s} = \max_{i\in\{\mathrm{L},\mathrm{R}\}}\left(\frac{M_i^2}{s}\right) \, ,
\end{equation}
over the two scaled invariant hemisphere masses $M_{\mathrm{L}/\mathrm{R}}$,
\begin{equation}
  \frac{M_{\mathrm{L}/\mathrm{R}}^2}{s} = \frac{1}{s}\left(\sum\limits_{j \in \mathcal{H}_{\mathrm{L}/\mathrm{R}}} p_j\right)^2 \, .
\end{equation}
The $\rho_\mathrm{H}$ and $\tau$ distributions are identical at \LO and as such vanish in the two-particle limit, $\rho_\mathrm{H} \to 0$. For three-particle events, the heavy-jet mass is bounded by $\rho_\mathrm{H} \leq \frac{1}{3}$.

The \Caesar parametrisation of the heavy-jet mass is $a=1$, $b=1$ and the remainder function entering Eq.~\eqref{eq:CAESAR_master} is given by \cite{Catani:1992ua,Catani:1991kz,Banfi:2001bz}
\begin{equation}
  \FFun_{\rho_\mathrm{H}}(L) = \frac{\exp\left(-\gammaE R'(L)\right)}{\Gamma\left(1+\frac{R'(L)}{2}\right)^2} \, .
\end{equation}

\paragraph{Jet broadening}
The total and wide jet broadening $B_\mathrm{T}$ and $B_\mathrm{W}$ are defined by \cite{Rakow:1981qn,Catani:1992jc}
\begin{equation}
  B_\mathrm{T} = B_\mathrm{L}+B_\mathrm{R} \,, \quad B_\mathrm{W} = \max(B_\mathrm{L},B_\mathrm{R}) \, ,
\end{equation}
in terms of the hemisphere broadening
\begin{equation}
  B_{\mathrm{L}/\mathrm{R}} = \frac{\sum\limits_{j\in \mathcal{H}_{\mathrm{L}/\mathrm{R}}}\mods{\vec{p}_j\times \vec{n}_T}}{2\sum\limits_j\mods{\vec{p}_j}} \, ,
\end{equation}
for the two hemispheres ${\cal{H}}_\mathrm{L}$, ${\cal{H}}_\mathrm{R}$ separated by the thrust vector $\vec{n}_T$. Both vanish in the two-jet limit, $B_\mathrm{T}\to 0$, $B_\mathrm{W} \to 0$, and are bounded from above by $B_\mathrm{T} = B_\mathrm{W} = \frac{1}{2\sqrt{3}}$ for three-parton configurations.

The \Caesar parametrisation of the jet broadenings is $a=1$, $b=0$, while the remainder functions entering Eq.~\eqref{eq:CAESAR_master} are given by \cite{Dokshitzer:1998kz,Banfi:2001bz,Herren:2022jej}
\begin{align}
  \FFun_{B_\mathrm{T}}(L) &= \frac{\exp\left(-\gammaE R'(L)\right)}{\Gamma(1+R'(L))}\left[2^{\frac{R'(L)}{2}}R'(L) \frac{\Gamma\left(\frac{R'(L)}{2}\right)}{\Gamma\left(2+\frac{R'(L)}{2}\right)}\, {}_2F_1\left(1,2,2+\frac{R'(L)}{2},-1\right)\right]^2 \, , \\
  \FFun_{B_\mathrm{W}}(L) &= \frac{\exp\left(-\gammaE R'(L)\right)}{\Gamma^2\left(1+\frac{R'(L)}{2}\right)}\left[2^{\frac{R'(L)}{2}}R'(L) \frac{\Gamma\left(\frac{R'(L)}{2}\right)}{\Gamma\left(2+\frac{R'(L)}{2}\right)}\, {}_2F_1\left(1,2,2+\frac{R'(L)}{2},-1\right)\right]^2 \, ,
\end{align}
with ${}_2F_1(a,b,c,z)$ the Gauss hypergeometric function.

\paragraph{Soft-drop thrust}
The soft-drop algorithm \cite{Larkoski:2014wba} recursively declusters the angular-ordered clustering sequence of the Cambridge--Aachen jet algorithm \cite{Dokshitzer:1997in,Wobisch:1998wt}.
In the $e^+e^-$ version, the Cambridge--Aachen algorithm is first applied the two hemispheres ${\cal{H}}_\mathrm{L}$ and ${\cal{H}}_\mathrm{R}$ separately until exactly one jet is left in each.
Subsequently, the soft-drop algorithm proceeds as follows:
\begin{enumerate}
\item decluster the last step of the clustering algorithm, splitting the jet into its constituents $i$ and $j$;
\item test the soft-drop criterion
  \begin{equation}
    \frac{\min\left(E_i,E_j\right)}{E_i+E_j} > z_\mathrm{cut} \theta_{ij}^\beta \, ,
  \end{equation}
  where $z_\mathrm{cut}$ and $\beta \geq 0$ are parameters of the soft-drop algorithm, $\theta_{ij}$ denotes the angle between the pseudojets $i$ and $j$, and $E_i$, $E_j$ denotes their energies;
\item if $i$ and $j$ fail the soft-drop criterion, the softer of the two pseudojets is discarded (groomed) and the algorithm starts over for the resulting harder pseudojet;
\item if instead $i$ and $j$ pass the soft-drop criterion, the algorithm terminates and the resulting hemisphere jet is the combination of $i$ and $j$.
\end{enumerate}
Based on this, we define soft-drop thrust as \cite{Baron:2018nfz,Marzani:2019evv}
\begin{equation}
  \tau_\mathrm{SD} = \frac{\sum\limits_{i\in \mathcal{E}_\mathrm{SD}}\mods{\vec p_i}}{\sum\limits_{i\in \mathcal{E}}\mods{\vec p_i}}\left(1 - \frac{\sum\limits_{i \in \mathcal{H}^\mathrm{SD}_\mathrm{L}}\mods{\vec n_\mathrm{L} \cdot \vec p_i} + \sum\limits_{i \in \mathcal{H}^\mathrm{SD}_\mathrm{R}}\mods{\vec n_\mathrm{R} \cdot \vec p_i}}{\sum\limits_{i\in\mathcal{E}_\mathrm{SD}}\mods{\vec p_i}} \right) \, ,
\end{equation}
where $\mathcal{E}$ denotes the whole event, $\mathcal{E}_\mathrm{SD}$ the soft-dropped event, and $\mathcal{H}_{\mathrm{L}/\mathrm{R}}^\mathrm{SD}$ the soft-dropped left and right hemisphere. The two vectors $\vec{n}_{\mathrm{L}/\mathrm{R}}$ denote the axis of the left- and right-hemisphere jet, respectively. 

For soft-drop thrust, the \Caesar parameters in Eq.~\eqref{eq:CAESAR_param} and the remainder function $\FFun$ entering Eq.~\eqref{eq:CAESAR_master}
are identical to the ungroomed case. In what follows we will specifically consider the three cases $\beta=0,1,2$ thereby keeping $\zcut=0.1$ fixed.

\section{Numerical validation in the soft limit}\label{sec:validation}

\begin{figure}[!ht]
  \centering
  \includegraphics[width=0.49\textwidth]{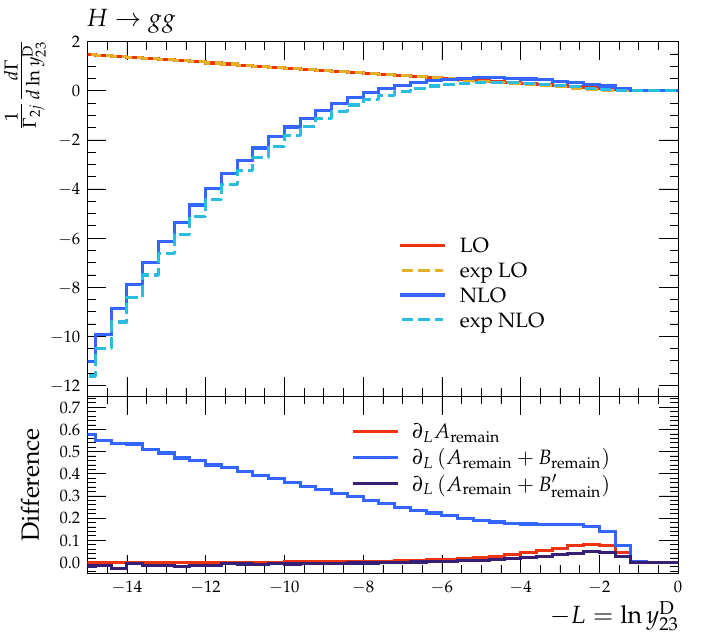}
  \includegraphics[width=0.49\textwidth]{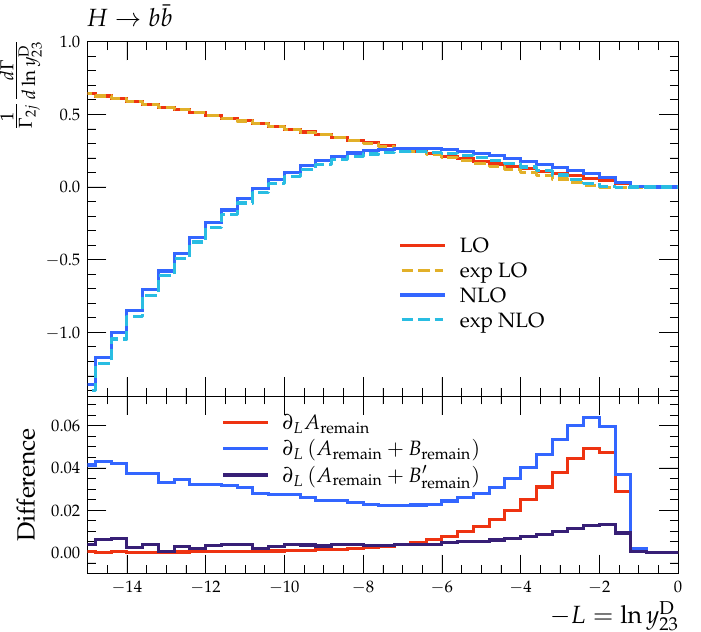}
  \caption{Comparison of the LO and NLO predictions with the corresponding
    expansion of the resummed result for the case of the Durham jet-resolution
    observable $y^\mathrm{D}_\mathrm{23}$ for the  $H\to gg$ (left) and $H\to b\bar{b}$ (right) Higgs decay modes. The lower panels show the difference between the fixed-order calculation and
    the expanded resummation, at order $\alphaS$ (red), $\alphaS^2$
    (blue), and when including the numerically extracted $\mathcal{C}_1$ coefficient
    (purple), see Eqs.~\eqref{eq:Aremain}, \eqref{eq:Bremain}, and \eqref{eq:Bpremain}
    for the definitions of $A_\text{remain}$, $B_\text{remain}$ and $B^\prime_\text{remain}$,
    respectively.}
  \label{fig:y23DiffPlots}
\end{figure}

We first cross-check the behaviour of our fixed-order predictions for the
two Higgs-boson decays in the soft limit against the ${\cal{O}}(\alphaS)$ and
${\cal{O}}(\alphaS^2)$ expansion of the resummed results. This provides a
crucial validation of both calculations. We have described the expected behaviour in
Sec.~\ref{sec:matching}. We begin by a detailed analysis of the results for the Durham
jet-resolution observable $y^\mathrm{D}_\mathrm{23}$. The validation for this observable is shown in
Fig.~\ref{fig:y23DiffPlots}, where we present the comparison for the $H\to
gg$ decay mode on the left-hand side and for the $H\to b\bar{b}$ case on the right-hand side. At leading order, we can
immediately see from the top panels that the exact three-parton tree-level matrix
element and the ${\cal{O}}(\alphaS)$ expansion of the resummation calculation, shown by the solid red and dashed orange lines, respectively,
approach one another in the soft limit, \emph{i.e.},\ for $y^\mathrm{D}_\mathrm{23}\to 0$.
Given that both results diverge towards $+\infty$ in this limit, it is more instructive
to consider the difference between the two results. 
This difference is shown in the lower panels of Fig.~\ref{fig:y23DiffPlots}
where we can confirm that it indeed
vanishes, \emph{i.e.}, $\partial_L A_\text{remain} \to 0$, in the soft limit (shown
in red). 

Moving on to the NLO prediction and the expansion of the resummed result up to
${\cal{O}}(\alphaS^2)$, shown as solid dark and dashed light blue lines in
Fig.~\ref{fig:y23DiffPlots}, we can first remark that they exhibit a qualitatively similar
behaviour as at LO, however, now diverging towards $-\infty$.
As expected, they do not exactly agree in the soft limit as the NLL resummation is missing
subleading logarithmic contributions at this order. By considering again the difference in
this kinematic limit, we observe a linearly rising difference, as can be expected from
Eq.~\eqref{eq:Bremain} (note we are here plotting the respective derivatives of the
quantities defined there). This artefact is much more prominent in the $H\to gg$ case,
where a linear rise from $0.3$ to $0.6$ can be observed in the range $L=8$ to $L=15$,
whereas the remainder in the $H\to b\bar{b}$ case only increases by $0.01$ over the same range.
Of course this has to be considered in relation to the fact that the whole distribution
in the second case is shifted towards larger logarithms, due to the smaller colour factors
involved in the case of Higgs decays to quarks. We hence argue that this behaviour is qualitatively
consistent with Casimir scaling. We have checked that the difference in $H\to b\bar{b}$ indeed
continues with a logarithmic rise, at least within the statistical uncertainties of
the numerical evaluation, until a value of $L=20$. Fluctuations do become large
in this regime and we hence limit the plot range as shown in the figure.

As argued before, our multiplicative matching scheme
incorporates an effective coefficient $A_0\equiv\mathcal{C}_1$, which captures the leading
part of this disagreement. This is illustrated in the lower panels by the purple curves,
where we subtract all the terms that are present to ${\cal{O}}(\alphaS^2)$ in our matched
predictions, \emph{cf.}\ Eq.~\eqref{eq:Bpremain}. We should stress that, while the remaining
difference indeed approaches a numerically small value, the two results still mismatch
by a constant term corresponding to an uncontrolled ${\cal{O}}(\alphaS^2 L)$ contribution
to $\Gamma_{gg}$ and $\Gamma_{b\bar{b}}$ that could only be captured by a NNLL accurate resummed
calculation. These observations hold for both cases, $H\to gg$ and $H\to
b\bar{b}$, in Fig.~\ref{fig:y23DiffPlots}.

\begin{figure}[h!]
  \centering
  \includegraphics[width=0.49\textwidth]{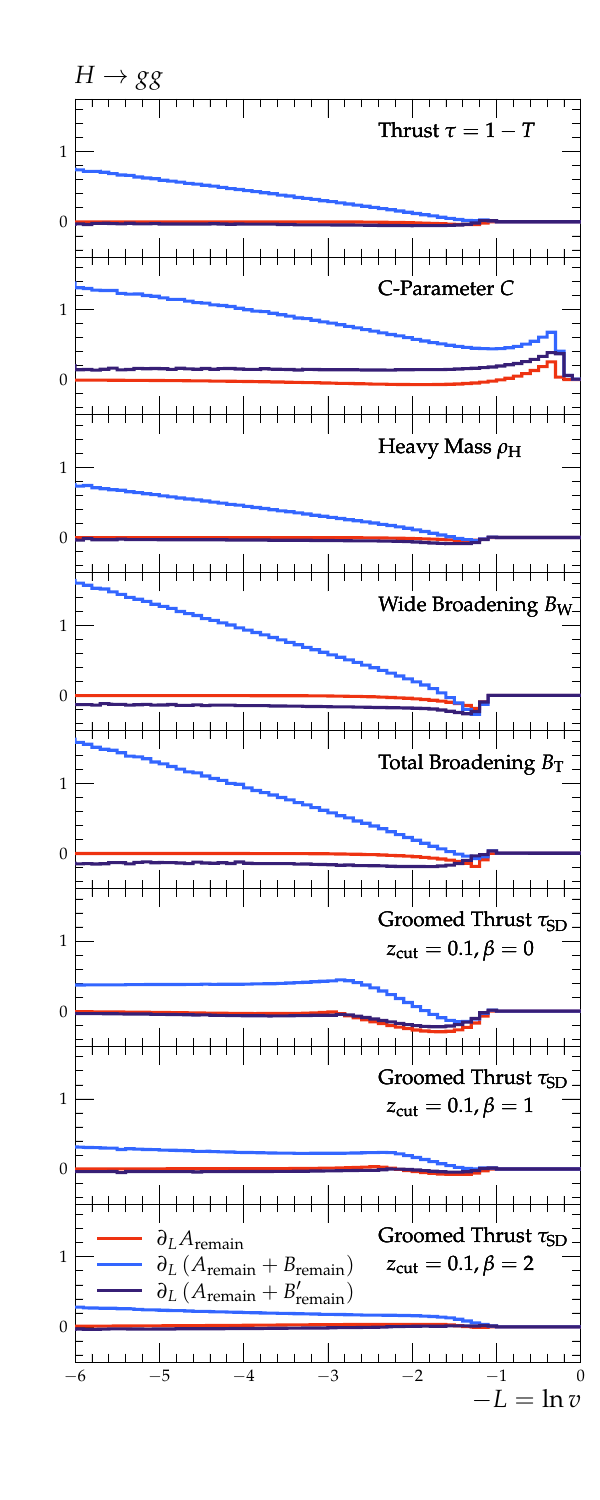}
  \includegraphics[width=0.49\textwidth]{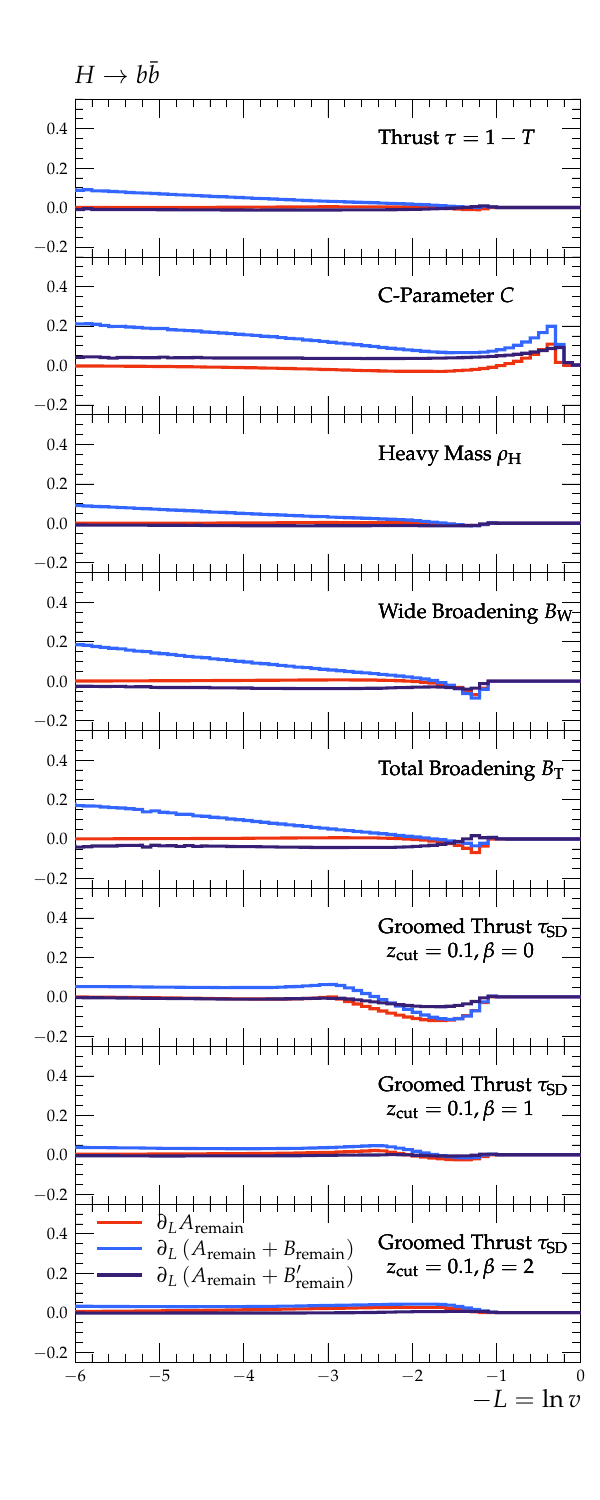}
  \caption{Residual differences between the LO and NLO calculation and the expanded
    resummation, at order $\alphaS$ (red), $\alphaS^2$ (blue), and when including the
    numerically extracted ${\cal{C}}_1$ coefficient (purple). For the various observables the
    left panel contains the results for the case $H\to gg$, while the right panel
    corresponds to $H\to b\bar{b}$.}
  \label{fig:DiffPlots}
\end{figure}

An overview of the same validation for the remaining observables can be
found in Fig.~\ref{fig:DiffPlots}. For the traditional (ungroomed) observables,
\emph{i.e.}, thrust $\tau$, $C$-parameter, scaled heavy-jet mass $\rho_\mathrm{H}$,
total and wide jet broadening $B_\mathrm{T}$ and $B_\mathrm{W}$, we find a similar picture as expected.
Again, the linear rise in the uncorrected remainder at NLO (shown in blue)
is much more prominent in the Higgs decaying to gluons decay mode, however for several observables, the linear
behaviour can unambiguously be identified also in the case where the Higgs boson decays into quarks.
Further, we note that for some observables like the $C$-parameter and the broadenings
we now indeed observe non-vanishing constant differences in $B^\prime_\text{remain}$
(purple lines). While these have a positive sign for the $C$-parameter in both Higgs-decay
categories, they exhibit a similar magnitude but with negative sign for both
$B_{\text{W}}$ and $B_{\text{T}}$.

We furthermore show results for soft-drop thrust for $\zcut=0.1$ and $\beta=0,1,2$
in Fig.~\ref{fig:DiffPlots}. In the groomed distributions, the logarithmic structure
is changed. For $\beta=0$, the leading logarithms in $\ln1/v$
are entirely replaced by logarithms of type $\ln 1/\zcut$, so that the remainder is
constant even without the inclusion of the additional terms in $B^\prime_\text{NLL}$.
In general, terms of this order are suppressed by powers of $\beta$. So for
$\beta>0$ the linearly rising behaviour of the remainder $B_\text{remain}$ is
restored, and again fixed by including the additional terms in
$B^\prime_\text{remain}$. 

\section{NLO+\NLLp  accurate predictions}\label{sec:resummed_predictions}

\begin{figure}[ht!]
  \centering
  \includegraphics[width=0.45\textwidth]{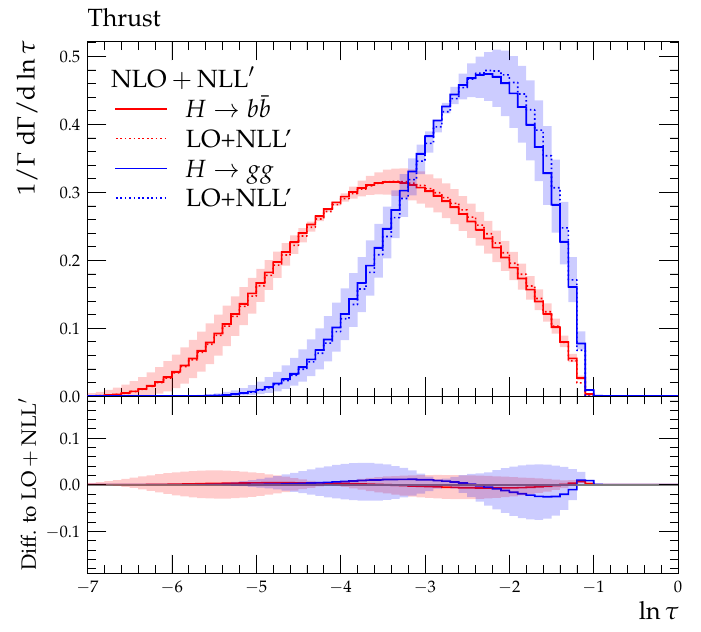}
  \includegraphics[width=0.45\textwidth]{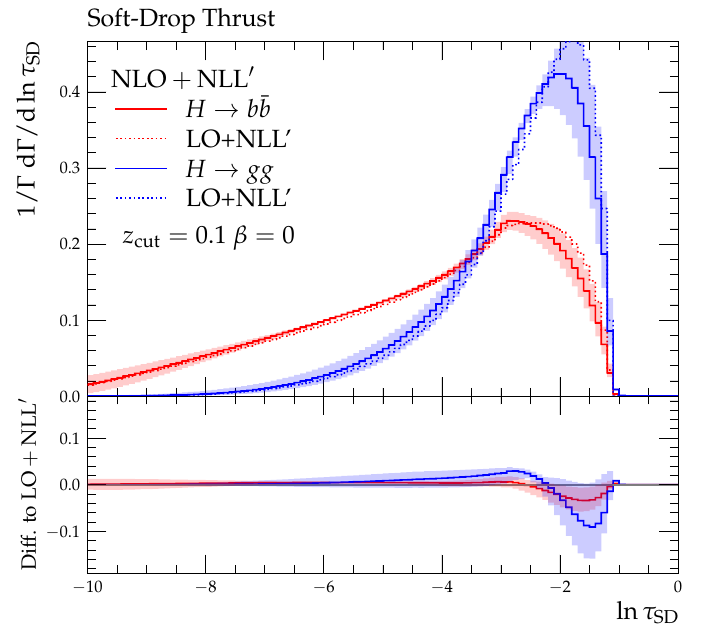}
  \caption{Matched NLO+NLL$^\prime$ (solid) and LO+NLL$^\prime$  (dashed) predictions for
    standard and soft-drop groomed thrust with $\beta=0$. The lower panels contain the
    difference between the respective \NLOpNLLp and \LOpNLLp results (see text). }
  \label{fig:ThrustFull}
\end{figure}

\begin{figure}[ht!]
  \centering
  \includegraphics[width=0.45\textwidth]{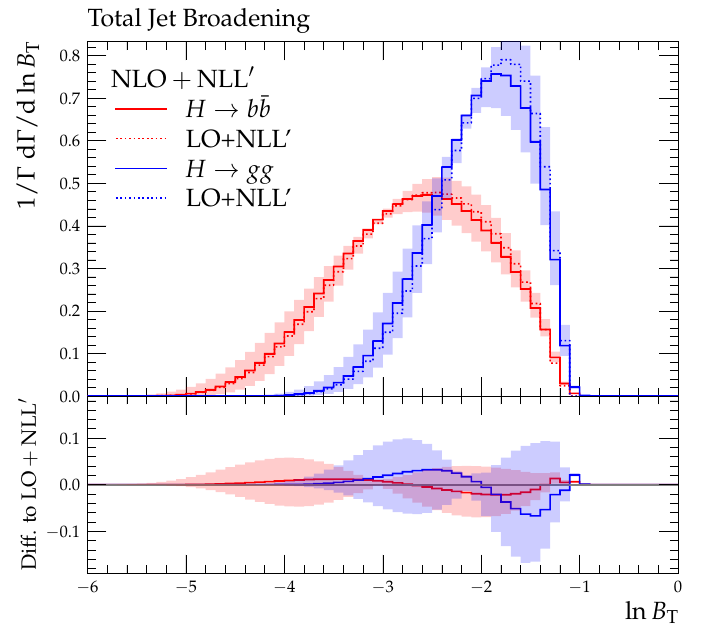}
  \caption{Matched NLO+NLL$^\prime$ (solid) and LO+NLL$^\prime$  (dashed) predictions for total broadening. The lower panel contains the difference between the respective \NLOpNLLp and \LOpNLLp results (see text).}
  \label{fig:BTFull}
\end{figure}

\begin{figure}[ht!]
  \centering
  \includegraphics[width=0.45\textwidth]{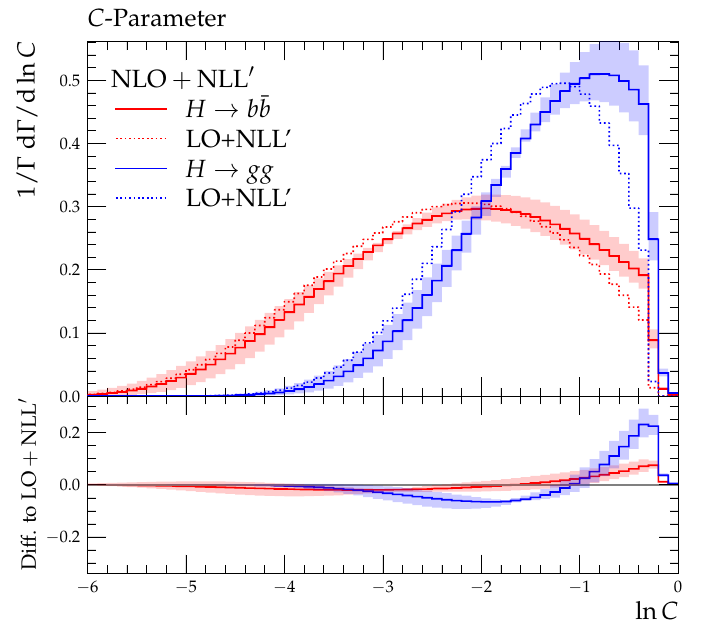}
  \includegraphics[width=0.45\textwidth]{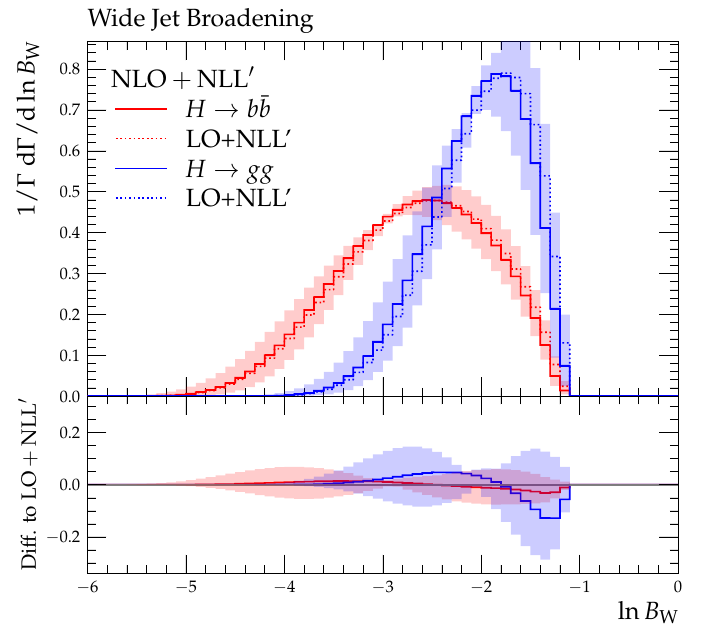}\\
  \includegraphics[width=0.45\textwidth]{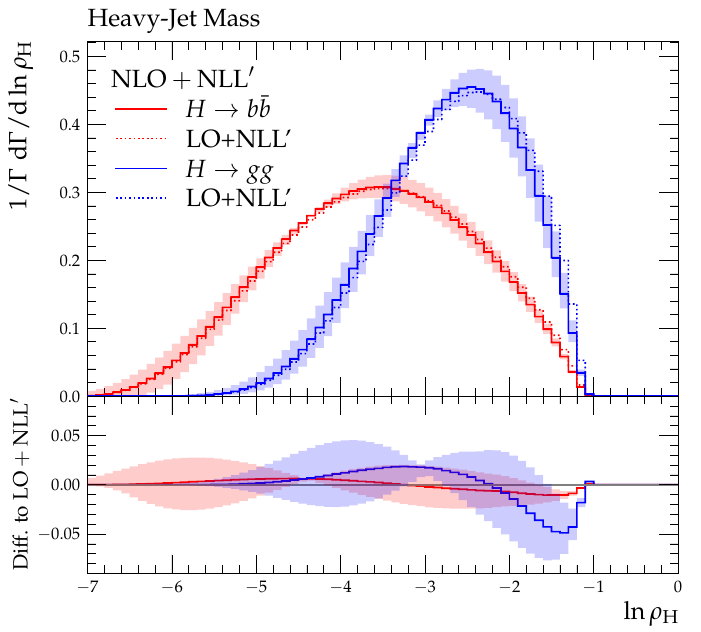}
  \includegraphics[width=0.45\textwidth]{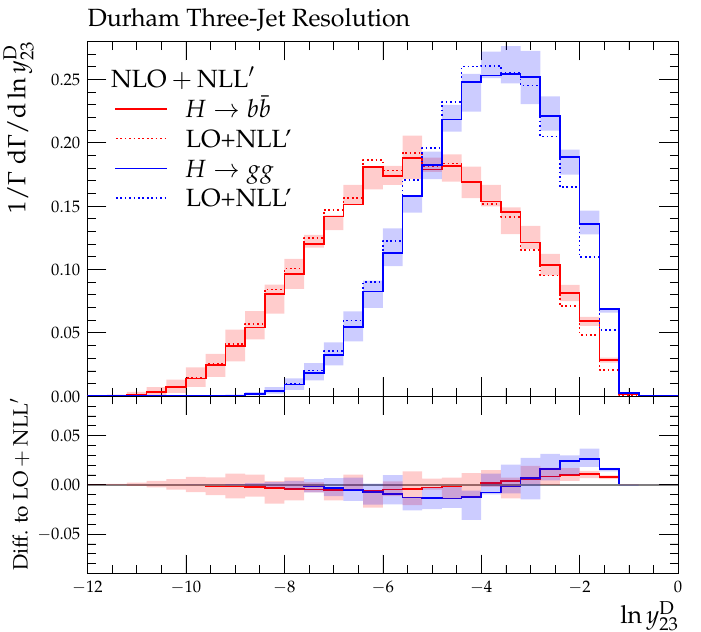}\\
  \includegraphics[width=0.45\textwidth]{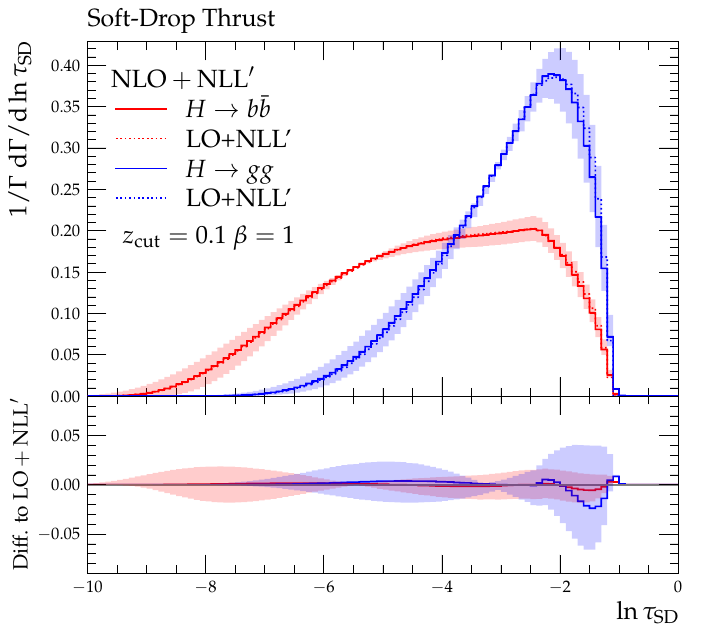}
  \includegraphics[width=0.45\textwidth]{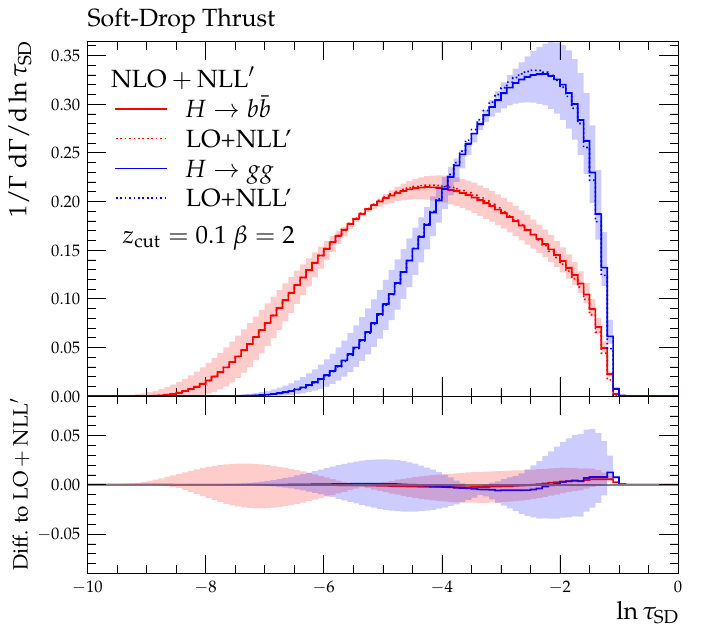}
  \caption{Matched NLO+NLL$^\prime$ (solid) and LO+NLL$^\prime$  (dashed) predictions for $C$-parameter and heavy-jet mass (top row), wide broadening and
    Durham jet resolution (middle row), and soft-drop groomed thrust with $\beta = 1,2$ (bottom row). The lower panels contain the difference between the respective
    \NLOpNLLp and \LOpNLLp results (see text).}
  \label{fig:ObservablesFull}
\end{figure}

We now move to the presentation of our final results, the matched distributions
for event-shape observables at NLO accuracy which after the multiplicative
combination with the NLL results ultimately achieve \NLOpNLLp precision.
In the following, we consider normalised distributions, \emph{i.e.}, all
histograms are divided by their respective cross section after matching
to the fixed-order result, therefore integrating to exactly one.
We thereby consider the following hadronic Higgs-decay observables: thrust, the $C$-parameter,
the heavy-hemisphere mass, total and wide jet broadening, and the Durham three-jet resolution
variable $y^\mathrm{D}_{23}$. Furthermore, we here present predictions for soft-drop
thrust~\cite{Marzani:2019evv} for three values of the soft-drop angular parameter $\beta$,
namely $\beta=0,1,2$. The results for those distributions are presented in
Figs.~\ref{fig:ThrustFull}--\ref{fig:ObservablesFull}.
Each of these figures, or sub-figures, consists of two panels, with the upper
panel showing the event-shape distribution and the lower panel presenting  
the difference between the respective \NLOpNLLp and \LOpNLLp results.

Unlike for the presentation of results in Fig.~\ref{fig:y23DiffPlots} and Fig.~\ref{fig:DiffPlots},
results for both Higgs decay categories, $H \to b\bar{b}$ and $H\to gg$, are presented together in the two panels of Figs.~\ref{fig:ThrustFull}--\ref{fig:ObservablesFull}.
The colour scheme in both panels is as follows: Results for the $H \to b\bar{b}$ decay mode are represented in red, while those associated to the $H \to gg$ decay mode are shown in blue. 
    
% scale variation for both
To estimate theoretical uncertainties, we
consider variations by factors of two around the central renormalisation-scale choice
$\muR=m_H$, thereby assuming $m_H = 125~\text{GeV}$, corresponding to the alternative choices
$\muR=2m_H$ and $\muR=m_H/2$. Similarly, we can vary the logarithm to be resummed to $L\to
\ln(x_L/v)$ while subtracting the additional induced NLL terms to retain our
logarithmic accuracy, see for example~\cite{Baberuxki:2019ifp} for details. We
likewise vary $x_L$ around our default $x_L=1$ to $x_L=2$
and $x_L=1/2$. Keeping the two variations separate, we overall obtain 5 results
and their spread is indicated in both panels, for the $H \to b\bar{b}$ decay
mode by the red and for the $H \to gg $ mode by the blue shaded bands, respectively.

% lower panels scale variation 
The impact of NLO corrections on the shape of the distributions can be seen when
  comparing the solid and dashed curves in the upper panels. Furthermore, they are illustrated
in the lower panels of Figs.~\ref{fig:ThrustFull}--\ref{fig:ObservablesFull}. The uncertainty
bands in these cases indicate the difference between the alternative scale choices in the
\NLOpNLLp and the central \LOpNLLp predictions. 

% observation general (size of NLO corrections in general) 
For the central scale choice, we observe the largest NLO corrections towards the
kinematical endpoints, \emph{i.e.}, for rather large observable values.
This is in line with expectations, because real-radiation corrections open up
the phase space in the multi-particle limit, located towards the right-hand side
of the figures. As such, event-shape observables are dominated by the real contribution
of the NLO corrections there.
Close inspection reveals that in these phase-space regions, the uncertainty bands
of the LO- and NLO-matched distributions do not overlap. This is consistent with
the observations made for the (unmatched) fixed-order predictions presented in
\cite{Coloretti:2022jcl}, \emph{i.e.}, that the LO distributions receive large NLO
corrections in both Higgs-decay channels.
In the limit of large logarithms, located on the left-hand side of the figures,
the differences between NLO- and LO-matched distributions are approaching zero, since
both distributions are smoothly vanishing in that region.
This is a requirement on physical distributions and ensured by our multiplicative
matching scheme, \emph{cf.} Eq.~\eqref{eq:GammaMatch}. 

Comparing the results obtained in both Higgs-decay categories, the following general statements hold.
As expected from Casimir scaling, the distributions in the $H\to gg$ decay mode are shifted
towards significantly higher values for all the event shapes considered here. Only for the
$\tau_{\text{SD}}$ distribution, this observation is broken to some extent, as soft-drop
grooming here significantly sculpts the $H\to b\bar{b}$ result. 

For soft-drop thrust there appears a transition point between the groomed and largely
ungroomed region at $\tau_\mathrm{SD} = \zcut 2^{\beta/2}$~\cite{Marzani:2019evv}. For
$\zcut=0.1$ as employed here, this is numerically close to the peak of the $H\to gg$ distribution,
\emph{e.g.}, for $\beta=0$ this corresponds to $\ln\tau_\text{SD}\approx -2.3$. For rather
strong grooming, in particular when using $\beta=0$, the quark distribution is significantly
suppressed below the transition point and in consequence develops a peak at larger values,
close to the maximum of the $H\to gg$ distribution. Interestingly,
while the distributions related to the Higgs to gluon decay mode approach zero much
earlier than those related to the Higgs to quark decay mode, as seen in the upper panels of the figures,
this is not necessarily the case for the difference between the respective LO- and NLO-matched distributions,
shown in the lower panels.

In Fig.~\ref{fig:ThrustFull} we present results for the
mass-like thrust $\tau$ with \Caesar parameters $a=1$, $b=1$ and its soft-drop groomed
variant $\tau_\mathrm{SD}$. As mentioned before, for both observables we present the distributions
containing results  for the gluonic $H\to gg$ decays as well as for the decays to $b$-quarks, in
the same plot.  In the groomed case, the difference above the transition point $\tau_\text{SD}\sim\zcut=0.1$
appears to be somewhat larger than in the equivalent region of the ungroomed
thrust. There appears a non-smooth feature around the transition point, \emph{i.e.},
$\ln\tau_\text{SD}\approx -2.3$, which in particular for the $H\to gg$ case is more
visible in the difference than in the actual distribution. Below this point the
difference to the leading-order matched distribution quickly decays.
In the ungroomed cases an oscillating behaviour is seen instead, with the difference
transitioning between positive and negative values, with vanishing absolute value.
We can observe in the two panels of Fig.~\ref{fig:ThrustFull} that 
the leading-order matched result appears to be within the error band of the \NLOpNLLp
result. The only exceptions are the points where the band boundaries cross each other,
as expected for normalised distributions. We conclude that, taking into account the
uncertainty estimates, the \LOpNLLp and \NLOpNLLp results are compatible.

In Fig.~\ref{fig:BTFull}, we show
our predictions for the transverse-momentum-like total broadening $B_\mathrm{T}$ with
\Caesar parameters $a=1, b=0$. The general hierarchy between the $H\to gg$ and
$H\to b\bar{b}$ cases is evident. We observe slightly larger corrections
in the difference between NLO- and LO-matched results in the lower panel. In this
case, the difference for gluon-initiated events indeed decays faster, just as
the overall distributions. Otherwise we observe a similar oscillating behaviour
as in the ungroomed thrust case. Close inspection of the hard phase-space region
on the far right-hand side of the plot again shows that the uncertainty bands
at LO and NLO do not overlap. As alluded to above, this is consistent with the
findings in \cite{Coloretti:2022jcl}. The distribution is dominated by the
real corrections at the hard kinematical endpoint.

In Fig.~\ref{fig:ObservablesFull}, we present corresponding results for the
remaining observables, \emph{i.e.}, for the $C$-parameter, wide jet broadening,
heavy-hemisphere mass, Durham three-jet resolution, and soft-drop thrust with $\beta=1,2$.
Qualitatively, we observe a similar impact of the NLO corrections as for thrust and
total jet broadening, \emph{cf.}~Figs.~\ref{fig:ThrustFull} and \ref{fig:BTFull}.
The largest NLO corrections in the difference can be seen for the $C$-parameter.
This is dominated by the Sudakov shoulder
effects~\cite{Catani:1997xc,Gehrmann-DeRidder:2007vsv,Bhattacharya:2022dtm,Bhattacharya:2023qet}
around the kinematic endpoints, \emph{i.e.}, the far-right side of the figure.
These effects emphasise that the respective leading-order results are not covered
by the NLO uncertainty band.
In this context, it is to be highlighted that all event-shape distributions shown here
are dominated by the \NLLp resummation in the soft region, $\ln v\lesssim -2$
and by the fixed-order calculation in the hard region, $\ln v\gtrsim -2$.
As expected, the heavy-jet mass behaves very similar to thrust; we note that they agree
for three-particle configurations.
Likewise, we observe a similar behaviour of the wide jet broadening $B_\mathrm{W}$
as for the closely related total jet broadening $B_\mathrm{T}$.

We note that the Durham three-jet resolution $y^\mathrm{D}_\mathrm{23}$ has a
similar scaling as the broadenings, however with $a=2,b=0$. This corresponds to
a factor two on the logarithm on the abscissa and we thus extend the plot range
correspondingly. The corrections from matching appear to be generally smaller in
this case than the others, in agreement with \cite{Coloretti:2022jcl}.
In particular we do not observe any marked features around the kinematical endpoint.

In the lowest row of Fig.~\ref{fig:ObservablesFull}, 
we finally show the results for soft-drop groomed thrust with two
different values of $\beta = 1$ and $\beta = 2$, corresponding to a less
aggressive grooming, compared to the case $\beta=0$. For the first case we observe
a transition behaviour with a peak at higher values in the $H\to b\bar{b}$ distribution,
moving the peak position closer to that of the $H\to gg$ case. For $\beta=2$, the grooming
is weak enough to allow for the development of the usual Sudakov peak, and we hence
observe a cleaner separation of the two distributions. The effects from NLO
matching compared to LO appear to be smaller again for these two groomed thrust
variants. There are also no easily identifiable features around the transition
point anymore in the difference shown in the lower panels. We only observe slight
differences towards the kinematical endpoint, which then vanish very fast with
increasing logarithm $L$, at least compared to the other cases we studied.

\section{Conclusions}\label{sec:conclusions}

Further scrutinising the Standard Model of particle physics through precision
measurements of Higgs-boson couplings forms a central task for experiments at
a possible future lepton collider. This in particular includes studies of
hadronic Higgs-boson decays that ultimately might provide access to an extraction
of the Higgs-gluon-gluon coupling. Event-shape observables offer a discriminatory power
between the various possible hadronic Higgs-boson decay channels, and in particular between the decay modes 
$H\to b\bar{b}$ and $H\to gg$, thereby being largely complementary to techniques
that instrument displaced vertices due to weak decays of flavourful hadrons. 

We have here presented resummed predictions at \NLOpNLLp accuracy for an 
extended set of three-jet event-shape observables for these two hadronic decay modes.
Besides the event-shape observables thrust, $C$-parameter, total and wide jet broadenings,
heavy-hemisphere mass, and the three-jet resolution in the Durham jet clustering algorithm,
we here also considered soft-drop groomed thrust.
Building on next-to-leading order calculations compiled with \eerad,
first presented in \cite{Coloretti:2022jcl}, we extended those to soft-drop
thrust and combined them with an all-orders resummation of next-to-leading logarithms
in the observables. To this end, we have used the implementation of the \Caesar
resummation formalism in the \Sherpa event-generator framework. Based on a multiplicative
matching scheme we were able to achieve \NLOpNLLp accuracy for the differential distributions
of the considered event-shape observables. We have carefully cross checked our fixed-order
and resummed calculations through a detailed comparison in softly and collinearly enhanced
kinematic configurations, \emph{i.e.}, for the case of very small observable values.

The results presented here extend the fixed-order predictions of \cite{Coloretti:2022jcl} to
phase-space regions in which event-shape observables receive large logarithmic corrections
from soft and collinear three-particle configurations. In the deep infrared region, at scales
of the order of $\Lambda_\mathrm{QCD}$, observables further receive important non-perturbative
corrections from the long-range behaviour of QCD. Hadronisation corrections are not included
in our predictions, but it can be anticipated that the impact of these can be reduced by
utilising grooming techniques, as done for the soft-drop thrust in the present work, see
\emph{e.g.}, \cite{Baron:2018nfz,Marzani:2019evv,Baron:2020xoi}.

In line with expectations, we find the impact of NLO corrections to be small over
the bulk of the logarithmic observable range. Considerable differences to the
leading-order-matched predictions are only found in the hard phase-space region. The largest
corrections were found for the $C$-parameter, amplified by the presence of large Sudakov-shoulder
effects, followed by the heavy-jet mass $\rho_\mathrm{H}$. The smallest NLO corrections are
present in the soft-drop thrust distributions, for all three values of $\beta$ studied.
For almost all event-shape observables considered here, we observe the expected behaviour
that distributions in $H\to gg$ decays peak at considerably larger observable values than
in $H\to b\bar b$ decays. This is only altered by grooming, which results in a shift of
the peak in $H\to b\bar b$ distributions towards larger observable values for $\beta = 0$
and $1$.
While we have here focussed on the six ``classical'' event shapes, including a groomed version
of thrust, our implementation straightforwardly extends to other (infrared-safe) observables,
so long as they fall in the category of being treatable by the \Caesar formalism. A particular
class of observables to mention are fractional moments of energy--energy correlators
$\mathrm{FC}_x$~\cite{Banfi:2004yd,Banfi:2018mcq,vanBeekveld:2023lsa} that recently were also
considered in the context of discriminating hadronic Higgs-boson decays~\cite{Knobbe:2023njd}.

Our work marks an important step towards precision calculations for Higgs-boson studies at future
lepton colliders, provides important benchmarks for the assessment
\cite{Hoeche:2017jsi,Dasgupta:2018nvj,Nagy:2020dvz,Nagy:2020rmk} and development
\cite{Dasgupta:2020fwr,Forshaw:2020wrq,Herren:2022jej,Assi:2023rbu,Hoche:2024dee,Hamilton:2023dwb,FerrarioRavasio:2023kyg,Preuss:2024vyu}
of parton-shower algorithms with higher logarithmic accuracy and can serve as theoretical input into jet-flavour tagging studies. 
To further improve the formal accuracy of predictions pertaining to three-jet event-shapes,
next-to-next-to-leading order (NNLO) corrections need to be evaluated, matching the
case of $\gamma^*/Z\to q\bar{q}$. To this end, all contributions, especially one- and
two-loop amplitudes, needed for NNLO corrections to three-jet observables in $H\to b\bar b$ decays
\cite{Gehrmann-DeRidder:2023uld,Ahmed:2014pka,Mondini:2019vub} and $H\to gg$ decays
\cite{Gehrmann-DeRidder:2023uld,Gehrmann:2011aa,Berger:2006sh,Badger:2006us,Badger:2007si,Glover:2008ffa,Badger:2009hw,Dixon:2009uk,Badger:2009vh}
are known in fully analytic form.
Similarly, the resummation should be lifted to the next-to-next-to-leading logarithmic level,
\emph{e.g.}, using the \Ares~\cite{Banfi:2014sua} framework.
When combined, NNLO+NNLL$^\prime$
accuracy could be achieved, resulting in a further reduction of theoretical uncertainties.
For certain ``simple'' observables, like thrust, $\text{NNLO}+\text{N}^3\text{LL}$ is within reach.
Relevant for the $H\to b\bar{b}$ channel, progress has been
reported on the inclusion of finite-mass effects, affecting the radiation pattern, in
resummed calculations~\cite{Fickinger:2016rfd, Gaggero:2022hmv, Ghira:2023bxr,Caletti:2023spr}.

\section*{Acknowledgements}

AG acknowledges support by the Swiss National Science Foundation (SNF) under contract 200021-197130.
SS acknowledges support from BMBF (05H21MGCAB) and funding by the Deutsche Forschungsgemeinschaft
(DFG, German Research Foundation) — project number 456104544.
The work of DR was supported by the STFC IPPP grant (ST/T001011/1).
Parts of the computations were carried out on the PLEIADES cluster at the
University of Wuppertal, supported by the Deutsche Forschungsgemeinschaft (DFG,
grant No. INST 218/78-1 FUGG) and the Bundesministerium f{\"u}r Bildung und
Forschung (BMBF).

%\appendix
%\section{some extra material}\label{app:formulas}

\bibliographystyle{amsunsrt_modp}
\bibliography{main}

\end{document}